\documentclass{PoS}
\usepackage{amsmath}
\usepackage{subfig}

\title{$K^0-\overline{K}^0$ mixing in the Standard Model from $N_f=2+1+1$ Twisted Mass Lattice QCD}

\ShortTitle{$K^0-\overline{K}^0$ mixing in the SM from $N_f=2+1+1$ tmLQCD}
\author {ETM Collaboration}

\author{\speaker{N.~Carrasco}, V.~Gim\'enez\\
        Dep. de F\'isica Te\`orica and IFIC, Universitat de Val\`encia-CSIC\\
        E-mail: \email{\{nuria.carrasco,vicente.gimenez\}@uv.es}}

\author{P.~Dimopoulos, R.~Frezzotti, D.~Palao, G.~C.~Rossi \\
        Dip. di Fisica, Universit\`a di Roma "Tor Vergata" and INFN-"Tor Vergata"\\
        E-mail: \email{\{dimopoulos,frezzotti,david.palao,rossig\}@roma2.infn.it}}

\author{V.~Lubicz\\
        Dip. di Fisica, Universit\`a Roma Tre and INFN-Roma Tre \\
        E-mail: \email{lubicz@fis.uniroma3.it}}

\author{M.~Papinutto\\
        Laboratoire de Physique Subatomique et de Cosmologie, UJF/CNRS-IN2P3/INPG\\
        E-mail: \email{Mauro.Papinutto@lpsc.in2p3.fr}}
        
\author{F.~Sanfilippo\\
        Dip. di Fisica, Universit\`a di Roma "La Sapienza" and INFN\\
        E-mail: \email{francesco.sanfilippo@roma1.infn.it}}

\author{S.~Simula\\
        INFN-Roma Tre \\
        E-mail: \email{simula@roma3.infn.it}}

\abstract{

We present preliminary results at $\beta=1.95$ ($a=0.077$ fm) on the first unquenched $N_f=2+1+1$ lattice 
computation of the $B_K$ parameter which controls the neutral kaon oscillations in the Standard Model.
Using $N_f = 2+1+1$ maximally twisted sea quarks and Osterwalder-Seiler valence quarks we achieve $O(a)$ 
improvement and a continuum-like renormalization pattern for the four-fermion operator. Our results 
are extrapolated/interpolated to the physical light/strange quark mass but not yet to the continuum limit.
The computation of the relevant renormalization constants is performed non perturbatively in the 
RI'-MOM scheme using dedicated simulations with $N_f=4$ degenerate sea quark flavours produced by the 
ETM collaboration. 

We get $B_K^{RGI}(a=0.077)\, =\, 0.747(18)$, which when compared to our previous unquenched
$N_{f}=2$ determination and most of the existing results, suggests a rather weak $B_K^{RGI}$ dependence on
the number of dynamical flavours. We are at the moment 
analysing lattice data at two additional $\beta$ values which will allow us to perform an extrapolation to the
continuum limit. 
 
}

\FullConference{ The XXIX International Symposium on Lattice Field Theory - Lattice 2011\\
July 10-16, 2011\\
Squaw Valley, Lake Tahoe, California}

\begin{document}

\section{Introduction}
The mixing $K^0-\overline{K}^0$  plays an important role in the understanding of the physics of CP-violation.
In the Standard model weak effective Hamiltonian (with three light
flavours) at the lowest order this oscillation is described only by
the operator
\begin{equation}
O^{\,\Delta S=2}\, =\, \frac{1}{4}\, [\overline{s}\, \gamma_{\mu}(1-\gamma_{5})\, d][\overline{s}\, \gamma^{\mu}(1-\gamma_{5})\, d]
\label{eq:Q1}
\end{equation}

The matrix element of the  $\Delta S=2$ effective hamiltonian can be factorized into a short distance contribution which can be
computed perturbatively and a nonperturbative long distance contribution containing the strong interaction effects.  These nonperturbative effects are described by the hadronic matrix elements of the renormalized four-fermion operator
\begin{equation}
\langle\overline{K}^{0}|\mathcal{H}_{eff}^{\Delta
S=2}|K^{0}\rangle=\frac{G_{F}^{2}M_{W}^{2}}{16\pi^{2}}{\displaystyle \left[{\displaystyle \sum_{l,m=u,c,t}C_{1}^{(l,m)}(\mu)\, V_{ls}^{*}V_{ld}V_{ms}^{*}V_{md}}\right]{\langle\overline{K}^{0}|\hat{O}^{\,\Delta S=2}(\mu)|K^{0}\rangle}}
\end{equation}

The $B_K$ parameter parametrizes the deviation of the hadronic element from the Vacuum Insertion Approximation (VIA)
\begin{equation}
\langle\overline{K}^{0}|\hat{O}^{\,\Delta S=2}(\mu)|K^{0}\rangle\, \equiv\, \langle\overline{K}^{0}|\hat{O}^{\,\Delta S=2}|K^{0}\rangle_{VIA}\; \hat{B}_{K}(\mu)\,=\, \frac{8}{3} f_{K}^{2} M_{K}^{2}\,\hat{B}_{K}(\mu)
\end{equation}

Combining the lattice computation of the $B_K$ parameter with the experimental value of $\epsilon_K$ one can constrain the values of the CKM matrix elements.

\section{Lattice setup}
In the gauge sector we use the Iwasaki action while the dynamical quarks have been regularized 
employing the twisted mass formalism  at maximal twist \cite{Frezzotti:2000nk} which provides automatic $\mathcal{O}(a)$ im\-pro\-ve\-ment \cite{Frezzotti:2003ni,Frezzotti:2003xj}. 
The fermionic action for the light doublet in the sea is given by
\begin{equation}
S_l^{sea,Mtm}={\displaystyle \sum_{x}\, \overline{\chi}_{l}(x)\, \left[D_{W}[U]+m_{0,l}+i\mu_{l}\gamma_{5}\tau_{3}\right]\, \chi_{l}(x) }
\end{equation}
where we follow the notation in \cite{Baron:2010bv}. In the heavy sector the sea quark action becomes
\begin{equation}
S_{h}^{sea,Mtm}\, =\, {\displaystyle \sum_{x}\, \overline{\chi}_{h}(x)\, \left[D_{W}[U]+m_{0,h}+i\mu_{\sigma}\gamma_{5}\tau_{1}+\mu_{\delta}\tau_{3}\right]}\, \mbox{\ensuremath{\chi}}_{h}(x)
\end{equation}

The $\mathcal{O}(a)$ improvement and a continuum-like renormalization pattern for the four-fermion operators can be achieved by introducing an Osterwalder-Seiler \cite{Osterwalder:1977pc} valence quark action allowing for a replica of the down $(d,d')$ and the strange $(s,s')$ quarks \cite{Frezzotti:2004wz}. The action for each OS valence flavour $\chi_f$ reads
\begin{equation}
S^{val,OS}_{f}\, =\, {\displaystyle \sum_{x}\, \overline{\chi}_{f}(x)}\, \left[D_{W}[U]+m_{0,f}+i\mu_{f}\gamma_{5}r_{f}\right]\, \chi_{f}(x)
\end{equation}
where the Wilson parameters should satisfy the relation $-r_s=r_d=r_{d'}=r_{s'}$.

In table \ref{tab:simulation-detail-bare} we give the details of the simulation of the bare $B_K$ parameter. For the inversions in the valence sector we used the stochastic method with propagator sources located at random timeslices in order to increase the statistical information \cite{Foster:1998vw, McNeile:2006bz}. 
\begin{center}
\begin{table}
\centering{} \begin{tabular}{|ccccc|}
\hline 
 & { $a\mu^{sea}$} & { $aM_{PS}^{ll}$ (MeV)} & { $L^{3}\times T$} & { \#confs}\tabularnewline
\hline 

 & { 0.0025} & { $\sim$270} & { $32^{3}\times64$} & { 144}\tabularnewline
{ $\beta$=1.95} & { 0.0035} & { $\sim$320} & { $32^{3}\times64$} & { 144}\tabularnewline
{ $(a\sim0.077\,\mbox{fm})$} & { 0.0055} & { $\sim$400} & { $32^{3}\times64$} & { 144}\tabularnewline
{ $a\mu_{\sigma}=0.135$ $a\mu_{\delta}=0.17$} & { 0.0075} & { $\sim$460} & { $32^{3}\times64$} & { 80}\tabularnewline
{ $a\mu^{val}=\{a\mu^{sea}$,0.0141,0.0180,0.0219\}} & { 0.0085} & { $\sim$490} & { $24^{3}\times48$} & { 244}\tabularnewline
\hline
\end{tabular}\caption{\label{tab:simulation-detail-bare}Simulation details for the bare $B_{K}$ parameter. The bare valence quark mass equal to $0.0141, 0.0180, 0.0219$ have been chosen to allow for a smooth interpolation around the physical strange quark mass. } 
\end{table} 
\end{center}
\vspace{-1cm}

The computation of the renormalization constants (RCs) for the relevant two- and four-fermion operators has been performed adopting the RI'-MOM scheme \cite{Martinelli:1994ty}. These RCs are computed by extrapolating to the chiral limit the RCs estimators measured at several quark mass values. 
We have performed dedicated runs with $N_f=4$ degenerate sea quarks in order to be able to estimate the chiral limit of the RCs.
In these $N_f=4$ simulations working at maximal twist would imply a considerable fine tuning effort due to the difficulties in determining $a m_{PCAC}$ near $a m_{PCAC}=0$. 
Instead, working out of maximal twist the stability of the simulations increases and the $\mathcal{O}(a)$ improvement of the RC estimators is achieved by averaging simulations 
with an equal value of the polar mass $M^{sea}$
but opposite value of $m_{PCAC}^{sea}$ and $\theta^{sea}$, where $\tan \theta^{sea}\, =\, \frac {Z_A\, m_{PCAC}^{sea}}{\mu^{sea}}$  \cite{Dimopoulos:2011wz}.
We label these ensembles as Ep/m where E=1,2... and p/m refers to the sign($\theta^{sea}$). In table \ref{tab:simulation-detail-RCs} we report the parameters of the ensembles analysed to compute the RCs.
\begin{center} 
\begin{table} 
\begin{tabular}{|cccccccc|} 
\hline 
ensemble & $a\mu^{sea}$ & $am_{PCAC}^{sea}$ & $aM^{sea}$ & $\theta^ {sea}$ & $a\mu^{val}$ & $am_{PCAC}^{val}$ &  \#confs \\
\hline
1m & 0.0085 & -0.04125(13) & 0.03308(10)& -1.3109(08) & set 1 & -0.02116(2) & 304 \\
1p & 0.0085 & +0.04249(13) & 0.03286(09)& +1.3091(08) & set 1 & +0.01947(19) & 304\\
\hline
7m & 0.0085 & -0.03530(13) & 0.02851(10)& -1.2681(10) & set 1 & -0.0216(2) & 400\\
7p & 0.0085 & +0.03608(11) & 0.02854(08)& +1.2683(09) & set 1 & +0.01947(19)& 400\\
\hline
8m & 0.0020 & -0.03627(11) & 0.02804(08) & -1.4994(02) & set 1 & -0.0216(2) & 400 \\
8p & 0.0020 & +0.03624(13) & 0.02743(10) & +1.4978(03) & set 1 & +0.01947(19) & 400\\
\hline
3m & 0.0180 & -0.0160(2) & 0.02191(09) & -0.6068(59) & set 2 & -0.0160(2) & 352\\
3p & 0.0180 & +0.0163(2) & 0.02183(09) & +0.6015(57) & set 2 & +0.0162(2) & 352\\
\hline
2m & 0.0085 & -0.02091(16) & 0.01815(11) & -1.0834(32) & set 1 & -0.0213(2) & 352\\
2p & 0.0085 & +0.0191(2) & 0.01692(13) & +1.0445(45) & set 1 & +0.01909(18) & 352\\
\hline
4m & 0.0085 & -0.01459(13) &  0.01404(08) & -0.9206(43) & set 2 & -0.01459(13) & 224 \\
4p & 0.0085 & +0.0151(2) & 0.01420(12) & +0.9289(64) & set 2 & +0.0151(2) & 224\\
\hline 
\end{tabular}\caption{\label{tab:simulation-detail-RCs}Details of the analysed ensembles for the RCs at 
$\beta=1.95$; set 1 $\, =\,$ \{$0.0085$, $0.0150$, $0.0203$, $0.0252$, $0.0298$\} and set 2 $\, =\,$ \{$0.0060$, 
$0.0085$, $0.0120$, $0.0150$, $0.0180$, $0.0203$, $0.0252$, $0.0298$\}. Here $M^{sea}\, =\, 
\sqrt{ (Z_A m_{PCAC}^{sea})^2 + (\mu^{sea})^2 }$, with a self-consistent $Z_A$-value.} 
\end{table} 
\end{center}

\section{The K-meson bag parameter}
The only four-fermion operator relevant for $B_K$ in the Standard Model is \footnote{Note that it contains the same physical information as \ref{eq:Q1}}  (see  \cite{Frezzotti:2004wz,Donini:1999sf} for notation)
\begin{equation}
{\rm Q}_1^{+}\, =\, O_{VV+AA}^{+}\, =\, [(\overline{s}\, \gamma_{\mu}\, d)(\overline{s}^{\prime}\, \gamma^{\mu}\, d^{\prime})\, +\, (\overline{s}\, \gamma_{\mu}\gamma_5\, d)(\overline{s}^{\prime}\, \gamma^{\mu}\gamma_5\, d^{\prime})]+[d\leftrightarrow d']
\end{equation}
It is multiplicatively renormalizable if the Wilson parameters appearing in the action of the OS fermions satisfy the relation $-r_s=r_d=r_{d'}=r_{s'}$ \cite{Frezzotti:2004wz}.

\begin{figure}
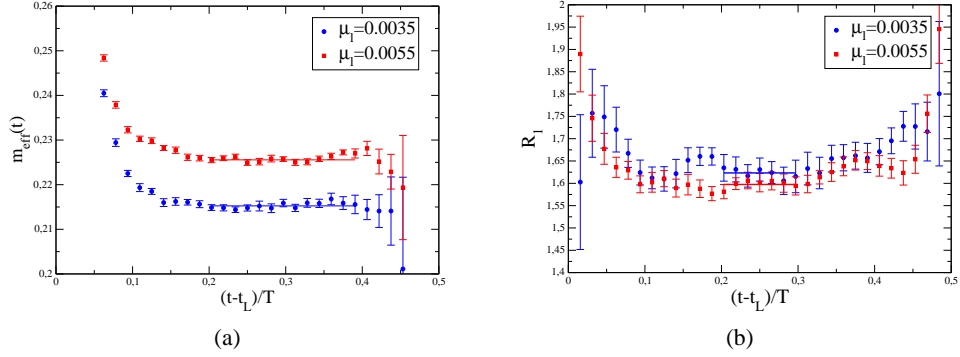

  \centering
  \subfloat[]{\label{fig:meff}\includegraphics[scale=0.17]{FIGURES/meffplateau.eps}}\hspace{1cm}
  \subfloat[]{\label{fig:plateau}\includegraphics[scale=0.17]{FIGURES/Rplateau.eps}}
  \caption{(a) Effective mass  of the pseudoscalar mesons made out of two mass-degenerate quarks regularized with opposite Wilson parameters (see \cite{Constantinou:2010qv} for definitions) (b) Quality of the ratio plateaus at fixed value of $\mu_h=0.0180$  }

\end{figure}

In order to estimate the $B_K$-parameter we compute a three point correlation function with an insertion of ${\rm Q}_{1}^{+}$ free to move in lattice time $t$, and two "$K$-meson walls" with pseudoescalar quantum numbers and with a fixed time separation between them $|t_L-t_R|=T/2$.
In this way, the lattice estimator for the bare $B_K$ parameter can be obtained from the ratio $R_1(t)$ for times $t_L\ll t\ll t_R$ 
\begin{equation}
R_1(t)\, =\, \frac{C^{(3)}_{\bar{K}{\rm Q}_1K}(t-t_L,t-t_R)}{C_{\bar{K}}^{(2)}(t-t_L)C_{K}^{(2)}(t-t_R)}\xrightarrow[t_L\ll t\ll t_R]{} \frac{8}{3}\, B_K
\end{equation}
where the two- and three-point correlation functions are defined as in \cite{Constantinou:2010qv}.
In figures \ref{fig:meff} and \ref{fig:plateau} the quality of the plateaus of the effective mass and the ratio $R_{1}$
are shown for two values of the light quark mass and one typical value of $\mu_h\, =\, 0.0180$.


\begin{figure}
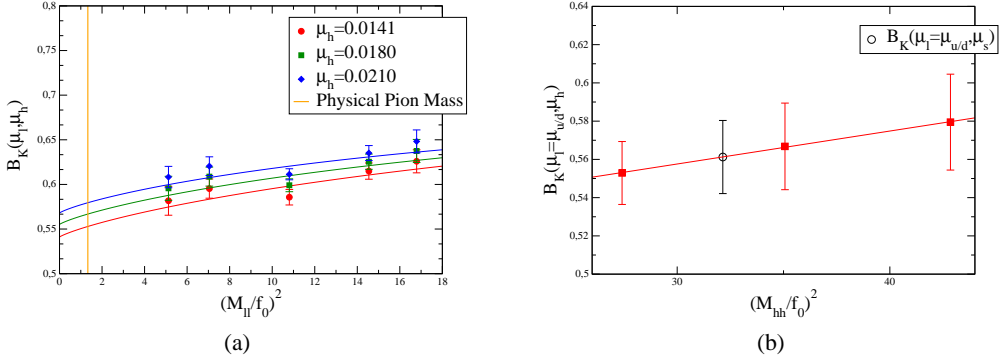

  \begin{center}
  \subfloat[]{\label{fig:extrap}\includegraphics[scale=0.17]{FIGURES/B1bare_extchiral.eps}}\hspace{1cm}
  \subfloat[]{\label{fig:interp}\includegraphics[scale=0.17]{FIGURES/B1bare_intstrange.eps}}
  \caption{\label{fig:tophyspoint} (a) Chiral fit according to the ansatz (3.3);  (b) Interpolation in the vicinity of the physical strange quark mass. }
  \end{center}
\end{figure}

Chiral extrapolations in the light quark mass are performed using SU(2) Partially Quenched Chiral Perturbation Theory at NLO
\cite{Allton:2008pn, Sharpe:1995qp}. We carry out three chiral fits, one for each simulated heavy mass, according to the fit ansatz  
\begin{equation}
B_{K}(M_{ll},M_{hh})\, =\, B_{K}^{'}(M_{hh})\; \left[1+b^{'}(M_{hh})\frac{(M_{ll})^{2}}{f_{0}^{2}}-\frac{(M_{ll})^{2}}{32\pi^{2}f_{0}^{2}}\log\frac{(M_{ll})^{2}}{16\pi^{2}f_{0}^{2}}\right]
\label{eq:SU2chfit}
\end{equation}
as we illustrate in figure \ref{fig:extrap}. 
The decay constant $f_0$ has been  obtained by analysing our data for the mass and decay constant of the $ll$ pseudoscalar meson  following the procedure of \cite{Baron:2010bv}.
Our result is $f_0=120.99(09)$ MeV. In order to estimate the systematic errors affecting this extrapolation in the $u/d$-quark mass, we also tried a first order polynomial fit.

Since we have obtained our lattice data for masses around the physical strange quark mass, we determine the 
bare $B_{K}$ parameter through an interpolation in $(M_{hh})^2$ to 
the physical mass $M_{hh}^2\, =\, 2\, M_K^2\, -\, M_{\pi}^2$ as is shown in figure \ref{fig:interp}.
 
Finally, the renormalized $B_K$ parameter is given by the relation \cite{Constantinou:2010qv},
\begin{equation}
\hat{B}_K\, =\, \frac{Z_{11}^{+}}{Z_A\, Z_V}\; B_K
\end{equation}
where $Z_{11}^{+}$, $Z_{A}$ and $Z_{V}$ are the RCs of the operators ${\rm Q}_{1}^{+}$, the axial and the vector currenty.
The relevant RCs both for the two- and four-fermion operators have been computed non-perturbatively in the RI'-MOM scheme following \cite{Martinelli:1994ty}.
 Our strategy to remove the inescapable $\mathcal{O}(a)$ discretization 
effects consists in performing a $\theta^{sea}$-average of the RCs estimators as described in \cite{Dimopoulos:2011wz}. 

\begin{figure}
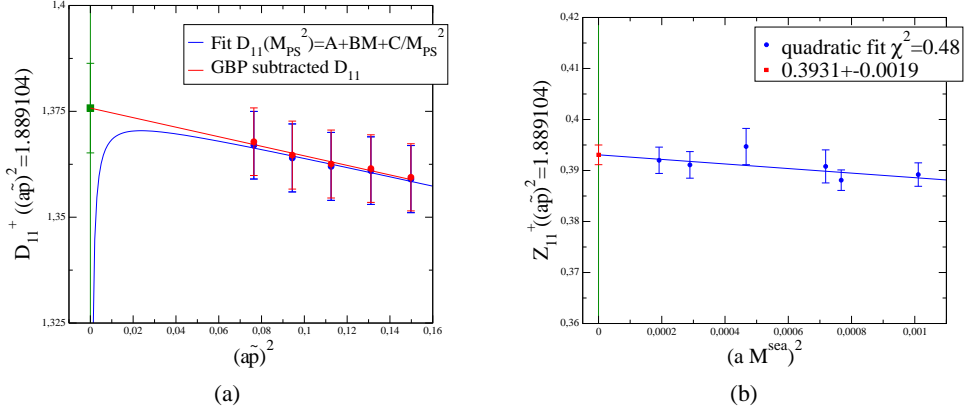

  \centering
  \subfloat[]{\label{fig:GPsubt}\includegraphics[scale=0.17]{FIGURES/D11PCp_B195_mu00085_pppm_12_2p_p128.eps}}\hspace{1cm}
  \subfloat[]{\label{fig:SeaExtr}\includegraphics[scale=0.17]{FIGURES/Z11_ext.eps}}
  \caption{(a) Goldstone boson pole substraction and valence chiral limit of $D_{11}$ for the ensemble 2p, $\mu^{sea}=0.0085$ and $(a\tilde{p})^2$=1.889104 ;  
  (b) Sea chiral limit extrapolation of $Z_{11}^{+}$ for $(a\tilde{p})^2\, =\, 1.889104$. }

\end{figure}

Using the notation of \cite{Donini:1999sf}, we compute the elements of the dynamical matrix $D_{ij}$ for each value of 
$a^2\tilde{p}^2\, \equiv\, \sum_{\nu}\, \sin^{2}(a p_{\nu})$, $\mu^{val}$ and Ep/m (and hence of $M^{sea}$). 
As described in \cite{Constantinou:2010qv}, the valence chiral limit extrapolation is safely determined with the help of the following fit ansatz
\begin{equation}
D_{ij}(\tilde{p}^2,\mbox{Ep/m}; \mu^{val})\, =\, D_{ij}(\tilde{p}^{2};\mbox{Ep/m})\, +\, A_{ij}(\tilde{p}^{2};\mbox{Ep/m})\, M^{val}\, +\, \frac{B_{ij}(\tilde{p}^{2};\mbox{Ep/m})}{(M_{PS}^{val})^2}
\end{equation}
at fixed values of $a^2\tilde{p}^2$ and Ep/m (and hence of $M^{sea}$). The last term takes into account the contribution of one Goldstone boson pole (GBP) (see Appendix A of \cite{Constantinou:2010qv} for details). 
In figure \ref{fig:GPsubt} we show the smooth dependence of $D_{11}$ on $(a M_{PS}^{val})^2$ for a representative value of the momentum even before the GBP substraction. Hence the one GBP contribution is strongly suppressed on the vertex $D_{11}$.
Having performed the GBP subtraction and the valence chiral extrapolation, 
the $\mathcal{O}(a)$ discretization effects are removed by averaging $D_{ij}(\tilde{p}^{2};\mbox{Ep/m}; \mu^{val}=0)$ over $\theta^{sea}$. 
By combining the valence chiral limit estimator of $D_{11}$ with the corresponding chiral limit of the quark field RC, $Z_q$, 
we obtain the intermediate quantities $Z_{11}^{+}(\tilde{p}^2,\mu^{val}=0,\mbox{Ep/m})$. Then, the sea 
chiral limit is taken at each fixed value of $(a\tilde{p})^2$ by fitting these intermediate estimators 
to a polynomial of first order in $(aM^{sea})^2$. We find that the dependence on the sea quark mass is 
very mild, as it can be seen from figure \ref{fig:SeaExtr}. In this way, we get the chiral limit RC estimators, 
$Z_{11}^{+}(\tilde{p}^2)$.

\begin{figure}
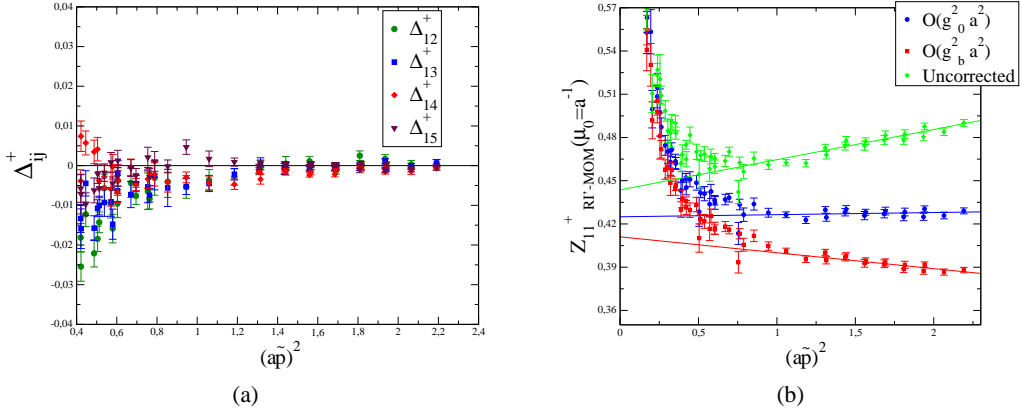

  \centering
  \subfloat[]{\label{fig:Delta}\includegraphics[scale=0.17]{FIGURES/Delta_uncor.eps}}\hspace{1cm}
  \subfloat[]{\label{fig:ZRIMOM}\includegraphics[scale=0.17]{FIGURES/Z11_RIMOM_compasubt.eps}}
  \caption{(a) Mixing coefficients in the valence and sea chiral limit ;   (b)  RI'-MOM computation of the multiplicative renormalization factor $Z_{11}^{+}$. }

\end{figure}

An important issue that arises in our analysis of the RCs is that wrong chirality mixings $\Delta_{ij}$ 
can affect the renormalization pattern of ${\rm Q}_1^{+}$ at order $a^2$ or higher \cite{Frezzotti:2004wz}. 
However, as it is clearly seen from figure \ref{fig:Delta}, these mixing coefficients are negligible within errors
in our data. 

Improved chiral limit RC estimators are obtained by analytically subtracting $\mathcal{O}(a^2g^2)$ 
discretization errors directly from the vertex $D_{ij}$, using the one-loop results in lattice perturbation
theory calculated in \cite{Constantinou:2010zs}. The effect of this perturbative correction is illustrated in figure 
\ref{fig:ZRIMOM} where we compare the RI'-MOM values for $Z_{11}^{+}$  at the reference scale 
$\mu_0=a^{-1}$ in three cases: uncorrected, corrected with a bare lattice coupling $g_{0}=6/\beta^2$ and 
with a boosted one $g_b^2\,\equiv\, g_{0}^2/\langle P\rangle$, where the average plaquette $\langle P\rangle$
is computed non perturbatively.

After bringing $Z_{11}^{+}(\tilde{p}^2)$ to a common reference scale $\tilde{p}^2\, =\,\mu_0^2\, =\, a^{-2}$ 
by employing the known NLO running formula \cite{Chetyrkin:1999pq}, we applied two different methods to remove the 
remaining $\mathcal{O}(a^2 \tilde{p}^2)$ discretization errors \cite{Constantinou:2010gr}. 
In the M1-method we perform a linear fit in $(a\, \tilde{p})^2$ of $Z_{11}^{+}(\mu_0^2)$ in 
the interval $(a\, \tilde{p})^2=[1.5:2.0]$ (see figure \ref{fig:ZRIMOM}). Alternatively, the M2-method 
consists in simply taking the weighted average of the RC in the momentum window $(a\, \tilde{p})^2=[1.8:2.0]$.

Combining four- and two-fermion RCs computed as in \cite{Dimopoulos:2011wz}, 
we obtain our final result for the RGI $B_K$ at $\beta=1.95$, corresponding to $a=0.077$ fm, by employing the estimates of the RCs from the procedure M1 and eq.(\ref{eq:SU2chfit}) for the extrapolation of our $B_K$ data to the u/d-quark physical point
$$ B_K^{RGI}(a=0.077)\, =\, 0.747(05)(17)[18] $$
where the first error is statistical, estimated from a boostrap analysis, and the second results from summing in quadrature several systematic uncertainties
(0.014 from the spread in  using eq.(\ref{eq:SU2chfit}) or a polynomial fit in the extrapolation to the physical point and 0.009 associated to the renormalization). The error quoted in brackets is the total error obtained by a sum in quadrature of the systematic and statistical ones.


\section{Acknowledgements}
The computer time for this project was made available to us by 
the PRACE Research Infrastructure resource JUGENE based in Germany at Forschungzentrum Juelich (FZJ)
and by BSC on MareNostrum in Barcelona (www.bsc.es).
N.C. and V.G. thank the MICINN (Spain) for partial support under Grant No. FPA2008-03373 and the Generalitat
Valenciana (Spain) for partial support under Grant No. GVPROMETEO2009-128. M. P. acknowledges financial support by a Marie Curie European Reintegration
Grant of the 7th European Community Framework Programme under contract
number PERG05-GA-2009-249309.

\end{document}